\documentclass[namedreferences]{kluwer}    

\usepackage{graphicx}

\newcommand{\Bf}{{\bf B}}
\newcommand{\ep}{{\bf e}_\phi}
\newcommand{\vf}{{\bf v}}
\newcommand{\pa}{\partial}
\newcommand{\Rs}{R_{\odot}}
\newcommand{\er}{\mbox{erf}}

\begin{document}                                                                                   

\begin{article}
 
\begin{opening}         

\title{Characteristics Of A Magnetic Buoyancy Driven Solar Dynamo Model} 
 
\author{Dibyendu \surname{Nandy}}

\runningauthor{Dibyendu Nandy}

\runningtitle{Magnetic Buoyancy And The Solar Dynamo}

\institute{Department of Physics, Indian Institute of Science, Bangalore 560012, India.}

\begin{ao}\\
Dibyendu Nandy,\\
Department of Physics,\\
Indian Institute of Science,\\
Bangalore 560012, India.\\
email: dandy@physics.iisc.ernet.in
\end{ao}
 
\begin{motto}
All models are wrong but some are useful. 
\rightline{George E. P. Box} 
\end{motto}

\begin{abstract}

We have developed a hybrid model of the solar dynamo on the lines of the
Babcock--Leighton idea that the poloidal field is generated at the surface
of the Sun from the decay of active regions. In this model magnetic buoyancy is handled with a realistic recipe - wherein toroidal flux is made to erupt from the overshoot layer wherever it exceeds a specified critical field ($10^5$ G). The erupted toroidal field is then acted upon by the $\alpha$ - effect near the surface to give rise to the poloidal field. In the first half of this paper we present a parameter space study of this model, to bring out similarities and differences between it and other well studied models of the past. In the second half of this paper we show that the mechanism of buoyant eruptions and the subsequent depletion of the toroidal field inside the overshoot layer, is capable of constraining the magnitude of the dynamo generated magnetic field there, although a global quenching mechanism is still required to ensure that the magnetic fields do not blow up. We also believe that a critical study of this mechanism may give us new information regarding the solar interior and end with an example, where we propose a method for estimating an upper limit of the diffusivity within the overshoot layer.                 
\end{abstract} 


\end{opening}

\section{Introduction}

Though there has not been a phenomenal change in our understanding of solar dynamo theory following the early seminal work of Parker, Steenbeck, Krause, and R\"adler (Parker 1955; Steenbeck, Krause, and R\"adler 1966, hereby the PSKR approach) and Babcock and Leighton (Babcock 1961; Leighton 1969, hereby the BL approach), each new study has in it's own way, contributed a little more to our understanding of the origin and evolution of the solar magnetic fields. Continuing in that line, we present here a study, that attempts to understand and quantify the effects of magnetic buoyancy on the solar dynamo.
 
Work on the solar dynamo is largely divided between the two approaches
(PSKR and BL) quoted above. The main difference between them being in the way the poloidal field generation is handled. Ones that follow the PSKR idea, invokes cyclonic turbulence in the interior of the solar convection zone (SCZ) to twist the toroidal field to generate the poloidal field (historically the $\alpha$-effect), while those following the BL idea, assumes that the poloidal field is generated from the decay of tilted active regions (resulting from the buoyant eruption of the toroidal field) on the solar surface. 
 
The toroidal field production process however, is the same in both these approaches and it is supposed to be generated due to the stretching of the poloidal field lines by differential rotation. We know that magnetic buoyancy is particularly destabilising in the SCZ (Parker 1975; Moreno-Insertis 1983) and therefore, the dynamo may not have enough time to amplify it to the very high values that the toroidal field seems to have. This led to the speculation that the dynamo generation of the toroidal field takes place in the overshoot layer beneath the SCZ (Spiegel and Weiss 1980; van Ballegooijen 1982; DeLuca and Gilman 1986; Choudhuri 1990) and with the helioseismic discovery of a strong radial shear layer (this region is also referred to as the tachocline and the overshoot layer is believed to be situated within this rotationally defined tachocline) in the differential rotation at the bottom of the SCZ, there remains little doubt that the toroidal field is indeed produced here within the overshoot layer.
   
Now, while the PSKR approach developed on strong foundations of mean field electrodynamics (Steenbeck, Krause, and R\"adler 1966; Moffatt 1978, Chap.\ 7; Parker 1979, \S18.3; Choudhuri 1998, \S16.5) and detailed models were worked out on the basis of this theory, the BL idea had to wait for quite sometime before detailed re-examination and {\it{recasting}} of the original ideas were attempted in the light of new results from flux tube rise simulations and helioseismology (Choudhuri, Sch\"ussler and Dikpati 1995; Durney 1995, 1996, 1997; Dikpati and Charbonneau 1999). 
 
In models based on the PSKR approach, magnetic buoyancy functions as a loss term in some treatments (DeLuca and Gilman 1986; Schmitt and Sch\"ussler 1989), while in others, a general upward flow due to magnetic buoyancy is included (Moss, Tuominen, and Brandenburg 1990a, 1990b). In BL models however, magnetic buoyancy not only removes flux from the overshoot layer but also contributes directly to the poloidal flux production procedure by transporting the strong toroidal field to the surface, thus playing a more significant role.
 
Amongst the recent BL models, with the exception of Durney's, the others worked with an $\alpha$-effect concentrated in a very thin layer near the solar surface - to capture the idea of the generation of poloidal field from the decay of active regions on the surface of the Sun. While Choudhuri, Sch\"ussler and Dikpati (1995) did not incorporate buoyancy, Dikpati and Charbonneau (1999) approximated magnetic buoyancy by making the source term for the generation of the poloidal field - proportional to the toroidal field at the bottom of the SCZ. Durney unlike the others, did away with the $\alpha$-effect altogether and generated the poloidal field by putting a double ring (analogous to a bipolar sunspot pair) on the surface at the same latitude - where he found the underlying toroidal field to be maximum, at specific intervals of time (for more details, see Durney 1997). 
  
So here we were at crossroads with the BL idea and the question naturally arose whether the {\it{$\alpha$-effect concentrated near the surface approximation}} and {\it{the double ring approximation}} are really very different from each other and if so, which one of these two methods, is a  more suitable expression of the BL idea? Nandy and Choudhuri (2001) [hereby Paper~I], showed that these two methods give qualitatively similar results. Paper~I also attempted to bridge the gap between the detailed mean field models and the more heuristic models based on BL ideas. 
   
A new way of handling magnetic buoyancy within a dynamo framework was introduced in Paper~I. With an algorithm wherein a certain fraction of the toroidal field in the overshoot layer was made to erupt to the top, at specific time intervals, wherever it exceeded a specified critical field $B_c$. The buoyant eruption was followed by a simultaneous depletion of the toroidal field within the overshoot layer. Although the two models presented in Durney (1997) and in Paper~I differ in the way they generate the poloidal field, perhaps Durney's method comes closest to the algorithm introduced in Paper~I in terms of a realistic recipe for handling buoyancy. Durney (1997) however, did not deplete the toroidal field in the overshoot layer (which he referred to as the GL) subsequent to eruptions, and depletion, as our present study shows, may have a profound influence on the magnitude and distribution of the dynamo generated fields at the bottom of the SCZ.
  
We, having demonstrated the viability of such a model - where a realistic algorithm for magnetic buoyancy works in tandem with a concentrated $\alpha$-effect near the top of the solar surface in Paper~I, now present a detailed analysis of such a model. In the present study we do not make any attempts at matching solar observations, rather, the emphasis is on understanding the effects of incorporating magnetic buoyancy. These hybrid models with mechanisms for handling buoyancy are, as of now, in their infancy and one has to do a critical study of such models to understand the physics underlying them. Section~2 details our model, in Section~3 we present our results. We end in Section~4 with a discussion highlighting the contribution of this model to our knowledge of the solar interior.
 
Note that in these introductory passages we have concentrated only on earlier work which is of direct relevance to our present study. Please refer to Dikpati and Charbonneau (1999) and Paper~I of this series (and references therein) for a more comprehensive review of the history of solar dynamo theory in general, and the BL approach in particular. 

\section{Buoyancy driven flux transport models}

\subsection{The model of Nandy and Choudhuri 2001 (Paper~I)}

Proceeding along the lines of the solar dynamo model with concentrated $\alpha$-effect presented in Paper~I, the evolution of the magnetic field can be expressed in terms of the vector potential $A$ [from which the poloidal field can be defined as $\Bf_p = \nabla \times (A \ep)$] and the toroidal field $B$, by the following two equations for the usual $\alpha \Omega$ dynamo:
\begin{equation}
\frac{\pa A}{\pa t} + \frac{1}{s}(\vf_p.\nabla)(s A)
= \eta \left( \nabla^2 - \frac{1}{s^2} \right) A + Q,
\end{equation} 
\begin{eqnarray}
\frac{\pa B}{\pa t} 
+ \frac{1}{r} \left[ \frac{\pa}{\pa r}
(r v_r B) + \frac{\pa}{\pa \theta}(v_{\theta} B) \right]
= \eta \left( \nabla^2 - \frac{1}{s^2} \right) B 
+ s(\Bf_p.\nabla)\Omega,
\end{eqnarray}
where $\eta$ is the coefficient of turbulent diffusion, $\Omega$ is the angular velocity, $\vf_p = v_r {\bf e}_r + v_{\theta} {\bf e}_{\theta}
$ is the meridional circulation and $s = r \sin \theta$.
 
We use a constant value of the turbulent diffusivity $\eta = 0.12 \times 10^8$ m$^2$ s$^{-1}$ for most of our calculations, unless otherwise stated. For the angular velocity $\Omega$, we use the same latitude-independent profile as in Paper~I. This expression for $\Omega$ is such that there is a strong radial shear concentrated in the tachocline below the SCZ, with a positive vertical gradient in the differential rotation - which corresponds to the helioseismologically determined profile at mid to low latitudes. For the meridional circulation also, we use the same profile as given in Paper~I, but with a higher value of ${v_0}$ = 10 m s$^{-1}$, for the maximum flow speed at mid-latitudes near the surface. This single cell flow (per meridional quadrant) is such that the flow is directed poleward near the surface and has a equatorward return flow near the bottom half of the SCZ.  
 
The term $Q$ on the right-hand side of Equation (1), is the source term for the generation of the poloidal field. Normally, the $\alpha \Omega$ dynamos are characterised by Equations (1) and (2) with $Q$ given by:
\begin{equation}
Q = \alpha B. 
\end{equation}

We use the following profile for the $\alpha$-coefficient:
\begin{eqnarray} 
\alpha =\frac{\alpha_0}{1 + {(B/B_0)}^2} \cos \theta \frac{1}{4}
\left[ 1 + \er \left(\frac{r - r_1}{d_1} \right) \right]
\times
\left[ 1 - \er \left(\frac{r - r_2}{d_2} \right) \right].
\end{eqnarray}
The parameters ($r_1$, $r_2$, $d_1$ and $d_2$) are adjusted to make sure that the $\alpha$ effect is concentrated near the surface of the Sun within $0.95 \Rs \leq r \leq \Rs$. We take $\alpha_0$ = 10 m s$^{-1}$ in most of the calculations in the present paper, which ensures that the dynamo is always supercritical and the solutions do not decay.   

The $\alpha$-quenching term $[1 + {(B/B_0)}^2]$ in the denominator of the source term in Equation (4) ensures that the poloidal field generation process gets suppressed when the erupted toroidal field has values close to, or higher than $B_0$. This quenching term is the only source of nonlinearity in models of such type and the dynamo generated magnetic fields scale according to the specified value of $B_0$. Results from flux tube simulations suggest that toroidal flux tubes, with magnitudes greater than $1.6 \times {10^5}$ G, emerge without any tilt on the solar surface (D'Silva and Choudhuri 1993; Fan, Fisher, and DeLuca 1993; Caligari, Moreno-Insertis, and Sch\"ussler 1995) and hence do not contribute to the generation of the poloidal field (resulting from the decay of tilted active regions). Following their work, we set ${B_0} = {10^5}$ G in Equation (4).   

Having specified the form of the $\alpha$-coefficient we now have to define an algorithm for buoyancy. Again drawing from our knowledge of flux tube simulations we know that toroidal flux tubes with values $\leq 0.6 \times {10^5}$ G rise parallel to the rotation axis, emerging at high latitudes with tilts that do not match Joy's Law (D'Silva and Choudhuri 1993; Fan, Fisher, and DeLuca 1993; Caligari, Moreno-Insertis, and Sch\"ussler 1995). Moreno-Insertis, Sch\"ussler, and Ferriz-Mas (1992) also showed that it is possible to store flux rings of strength $\leq {10^5}$ G within the overshoot region, while flux tubes greater in strength escape out. 

All this suggests that there should be a critical field beyond which flux tubes become buoyant and emerge radially to give rise to the sunspots and that the value of this critical field (which we shall denote as $B_c$) should be around ${10^5}$ G. Keeping these ideas in mind we have formulated a {\it{recipe}} for incorporating buoyancy within a dynamo framework in Paper~I. We summarise the salient features of this {\it{recipe}} below.
 
At intervals of time $\tau$ we check if the toroidal field $B$ has exceeded the specified critical value $B_c$ (${10^5}$ G), anywhere within the overshoot layer. Wherever the toroidal field exceeds $B_c$, a certain fraction $f_b$ of it is made to erupt radially, to the top layer near the surface - where the $\alpha$-effect is concentrated. The erupted toroidal flux $f'B$ (where $f' = {f_b}[R_i/R_f]$; $R_i$ is the radius at the bottom
from where the eruption takes place and $R_f$ is the radius near the
surface where the flux is deposited)\footnote{This takes into account the greater latitudinal extent of a grid size near the surface as compared to that at the bottom.} is added to the previously existing toroidal field near the surface (at the same latitude from where the eruption occurred), while the amount ${f_b}B$, is subtracted from the point inside the overshoot layer - which was the source of the eruption. Thus we make sure that flux is conserved in this procedure. We fix the time interval between successive eruption $\tau$ at $8.8 \times 10^5$ s, which allows for an order of a thousand eruptions in a complete dynamo period ($T_d$). Also, for most of the calculations presented in the results section, we use the value $f_b = 0.05$ for the control parameter (we call the fraction of the erupted field as the control parameter following Paper~I, because, this parameter controls the strength of magnetic buoyancy). This value corresponds to the {\it{buoyancy saturated regime}} of the dynamo (where buoyancy is quite strong and buoyant flux transport plays the main role in transporting flux from the bottom of the SCZ to the top).  
 
With this algorithm for buoyancy we solve Equations (1) and (2) in the northern quadrant of the convection (i.e.\ within $R_b = 0.7 \Rs \leq r \leq \Rs$, $0 \leq \theta \leq \pi/2$). For a description of the boundary conditions refer to Paper~I.  
           
\subsection{A buoyancy algorithm where the poloidal source term is proportional to the toroidal field inside the overshoot layer}

For reasons that will become clear as we go on, we felt it may be prudent to compare the results of the model defined by the buoyancy $\it{recipe}$ in Section~2.1 (hereby Model~1) with that of another model, the details of which follows.
  
This model (hereby Model~2) is similar in all respects to Model~1, except that instead of using the previous buoyancy algorithm (outlined in Section ~2.1), here we use a source term for the generation of the poloidal field which is proportional to the toroidal field at the bottom ($B_{\rm bot}$), inside the overshoot layer. This kind of a source term to model the decay of active regions was introduced by Choudhuri and Dikpati (1999) and was followed later by Charbonneau and Dikpati (1999) - who used this within a dynamo framework. For a motivation on its formulation see the above cited papers. Thus we replace Equation (4) and the buoyancy algorithm and work with Equation (1) and (2) along with a form of Equation (3) given by:

\pagebreak   

\begin{eqnarray} 
Q(r,\theta) =\frac{{\alpha_0}[B(r, \theta) + f_r B_{\rm bot} (\theta)]}
{1 + [\{B(r, \theta) + f_r B_{\rm bot} (\theta)\}/B_0]^2} 
\cos \theta \frac{1}{4}
\left[ 1 + \er \left(\frac{r - r_1}{d_1} \right) \right]
\nonumber
\\
\times
\left[ 1 - \er \left(\frac{r - r_2}{d_2} \right) \right].
\end{eqnarray} 
A notable difference between the source term used in Dikpati and Charbonneau (1999) and the one above is the inclusion of a term $f_r$ here, a parameter which controls how effective buoyancy is. Also the quenching expression here is slightly different; which instead of only accounting for the buoyant rise of field also incorporates the local field which is present near the surface already.

Equation (1) and (2) along with (5), with similar forms of meridional flow and the rotation profile, as described in Section~2.1, constitute our Model~2. We keep the values of $\alpha_0$ and $B_0$ the same as that of Model~1 to facilitate comparison.

\section{Results}

We have divided this section into two parts. The first part presents a parameter space study of Model~1 ending with a comparison with Model~2. The second part studies the effect of magnetic buoyancy on the dynamo generated magnetic fields and also discusses a novel way of calculating an upper limit of the diffusivity within the overshoot layer (this second part of the study is limited to Model~1).

\subsection{Variation of basic parameters}

There are quite a few parameters which are used as inputs in our model. Most of them have already been specified in Section~2. Notable amongst these and which also have featured prominently in the past literature on other solar dynamo models are; the amplitude of the meridional flow speed $v_0$, the amplitude of the source term for the generation of the poloidal field $\alpha_0$ and the diffusivity $\eta$. Parameters unique to the buoyancy driven model we are studying are; the critical field $B_c$, the time between eruption $\tau$ and the fraction of the erupted field $f_b$ (Model~1) or $f_r$ (Model~2). The emphasis is on studying the influence of these parameters on the dynamo period $T_d$, a quantity which typifies any dynamo model with a particular set of parameters (and also measures the efficiency of any cycle). Wherever appropriate, we also comment on the effect of varying these parameters on the dynamo generated magnetic fields. While varying any one parameter, we keep the other parameters constant at the values already mentioned in Section~2. 
 
\begin{figure}[t]
\centerline{\includegraphics[height=6.5cm, width=10cm]{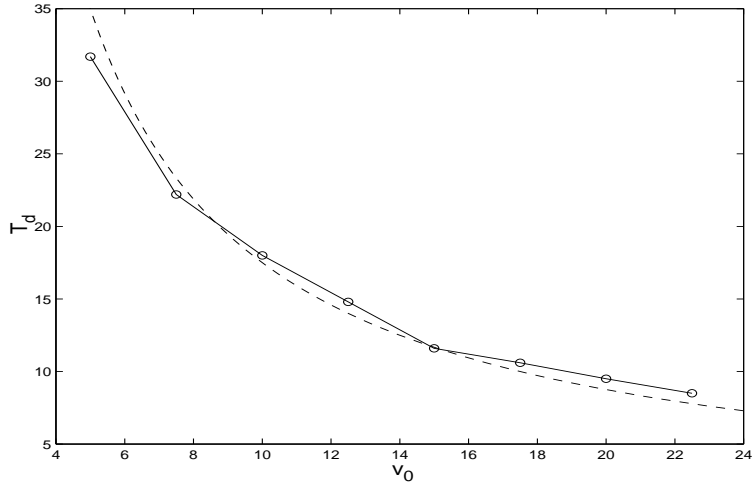}}
\caption{Variation of the dynamo period $T_d$ with the amplitude of the meridional flow speed $v_0$. $T_d$ is in years and $v_0$ is in m s$^{-1}$. The solid line connects the data points denoted by circles, while the dashed line depicts a C/x (C = a constant) behavior for comparison. Other parameters are; $\alpha_0$ = 10 m s$^{-1}$, $\eta = 0.12 \times 10^8$ m$^2$ s$^{-1}$, $B_0 = 10^5$ G, $B_c = 10^5$ G, $f_b = 0.05$.}
\end{figure} 

We start by presenting a plot of the variation of $T_d$ with $v_0$ for Model~1 in Figure~1, other parameters being the same as mentioned in Section~2 and with $f_b = 0.05$. We see that $T_d$ decreases with increasing $v_0$ and the dependence is almost $v_0^{-1}$ (as is evident from a comparison of the solid line connecting the data points and the dashed $v_0^{-1}$ line). In most PSKR models (many of which were worked out when the existence and role of meridional circulation was still not clear), turbulent diffusivity acted as the bridge between the source regions of the toroidal and the poloidal field - which often overlapped in these models.  Contrary to that, in BL models, meridional circulation plays an important role in transporting flux between the two source regions of toroidal and poloidal field production (from the bottom of the SCZ to the top near the equator and vice-versa near the pole). Therefore an increase in $v_0$ would mean faster flux transport - a more efficient cycle and hence an inverse dependence of $T_d$ on $v_0$. 
 
In our Model~1 however, magnetic buoyancy is also involved in the flux transport process - over a much wider extent in latitude and also at a much faster rate. One would have expected then that the dependence of $T_d$ on $v_0$ would be much less pronounced in this case. Nonetheless, we still find a drastic dependence of $T_d$ on $v_0$. It seems that although buoyancy may be more efficient in transporting flux to the surface from the overshoot layer, the crucial factor in completing the chain turns out to be the transport of the poloidal field from the surface (near the poles) to the bottom of the SCZ for the regeneration of the toroidal field - a process which can only be carried out by the meridional down-flow near the poles. Hence, it turns out that the dynamo period is critically dependent on the meridional flow even for buoyancy driven models. If we make $v_0 < 5.0$ m s$^{-1}$, the dynamo wave at the bottom of the SCZ starts propagating poleward, that is the dynamo is no longer advection-dominated.

\begin{figure}[t]
\centerline{\includegraphics[height=6.5cm, width=10cm]{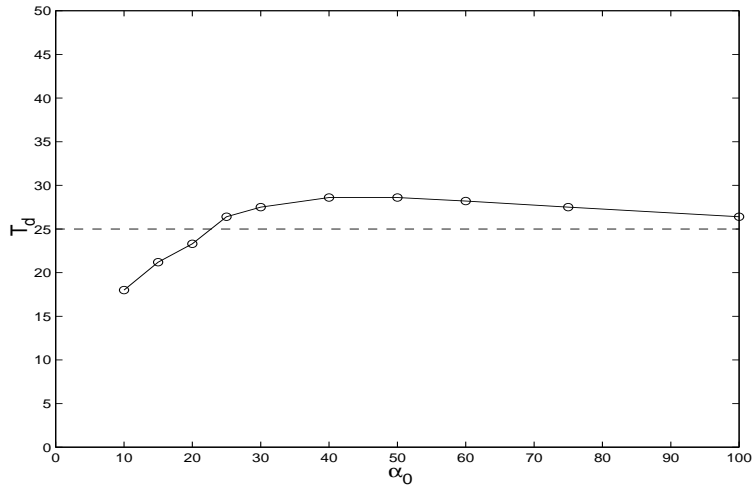}}
\caption{Variation of $T_d$ (in years) with the amplitude of the source coefficient $\alpha_0$ (in m s$^{-1}$). The solid line connects the data points and the dashed line shows a y = constant behavior, for comparison. Other parameters are; $v_0$ = 10 m s$^{-1}$, $\eta = 0.12 \times 10^8$ m$^2$ s$^{-1}$, $B_0 = 10^5$ G, $B_c = 10^5$ G, $f_b = 0.05$.}
\end{figure}

Figure~2 shows the variation of the dynamo period on changing the amplitude of the $\alpha$-coefficient for Model~1. We see that a varying $\alpha_0$ does not have much influence on $T_d$. This is fortunate for us because $\alpha_0$ essentially is the strength of a phenomenological source term, which represents the decay of tilted active regions to produce the poloidal field. Within a buoyancy driven dynamo framework then, one would like $T_d$ to be influenced more by the buoyancy mechanism (parameters which control the buoyant flux transport), rather than the amplitude of the phenomenological source coefficient. Moreover a reliable estimate of $\alpha_0$ is a formidable task, specially at the non-linear quenched regime, whether it be the BL approach motivated $\alpha$ or the PSKR approach $\alpha$ (Pouquet, Frisch, and Leorat 1976; Brandenburg and Schmitt 1998). Dynamo periods of most models based on the BL approach are actually found to be rather independent of the source coefficient. In contrast to this, dynamo periods of models based on the PSKR approach are significantly dependent on the strength of the $\alpha$-effect. 

If we make $\alpha_0 < 10.0$ m s$^{-1}$, the dynamo becomes sub-critical and we get decaying solutions. Also, the poloidal field near the pole increases rapidly if $\alpha_0$ is increased. A generic problem of this kind of BL models anyway, is the existence of high fields near the pole. Therefore it may be a good idea to work with a low value of $\alpha_0$, keeping the dynamo just about super-critical.      

The Dikpati and Charbonneau (1999) model is sufficiently different from the model that we are working with at present (for example they also incorporate the latitudinal dependence in the differential rotation and a depth-dependent diffusivity in their model). Given that they had reported a drastically different dynamo period dependence on diffusivity, we wanted to explore how a poloidal field source term similar to their buoyancy {\it{recipe}} behaves, within the framework of our model and hence, we constructed Model~2. We now present some results to facilitate comparison between our Model~1 and Model~2. 

\begin{figure}[t]
\centerline{\includegraphics[height=6.5cm, width=10cm]{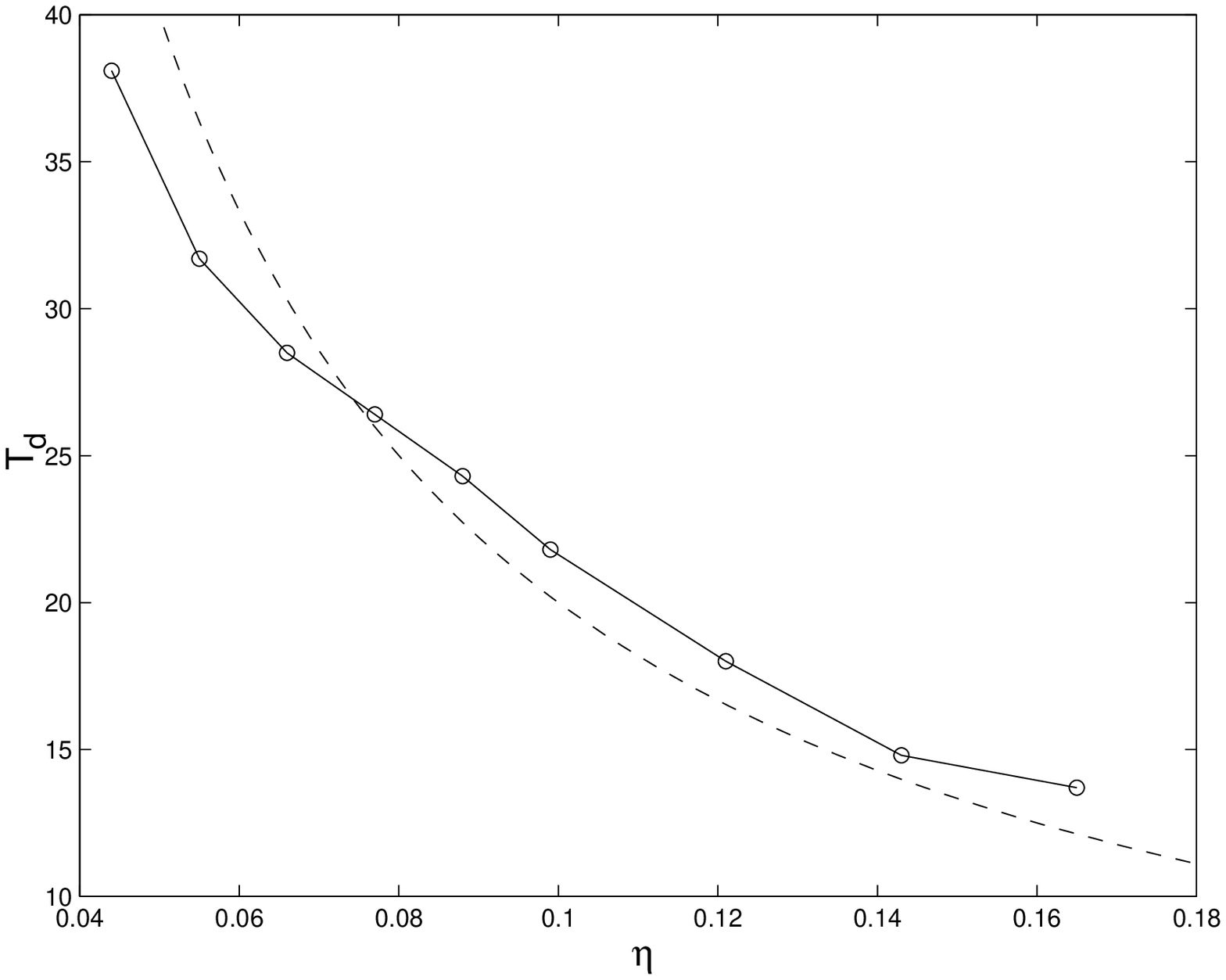}}
\centerline{\includegraphics[height=6.5cm, width=10cm]{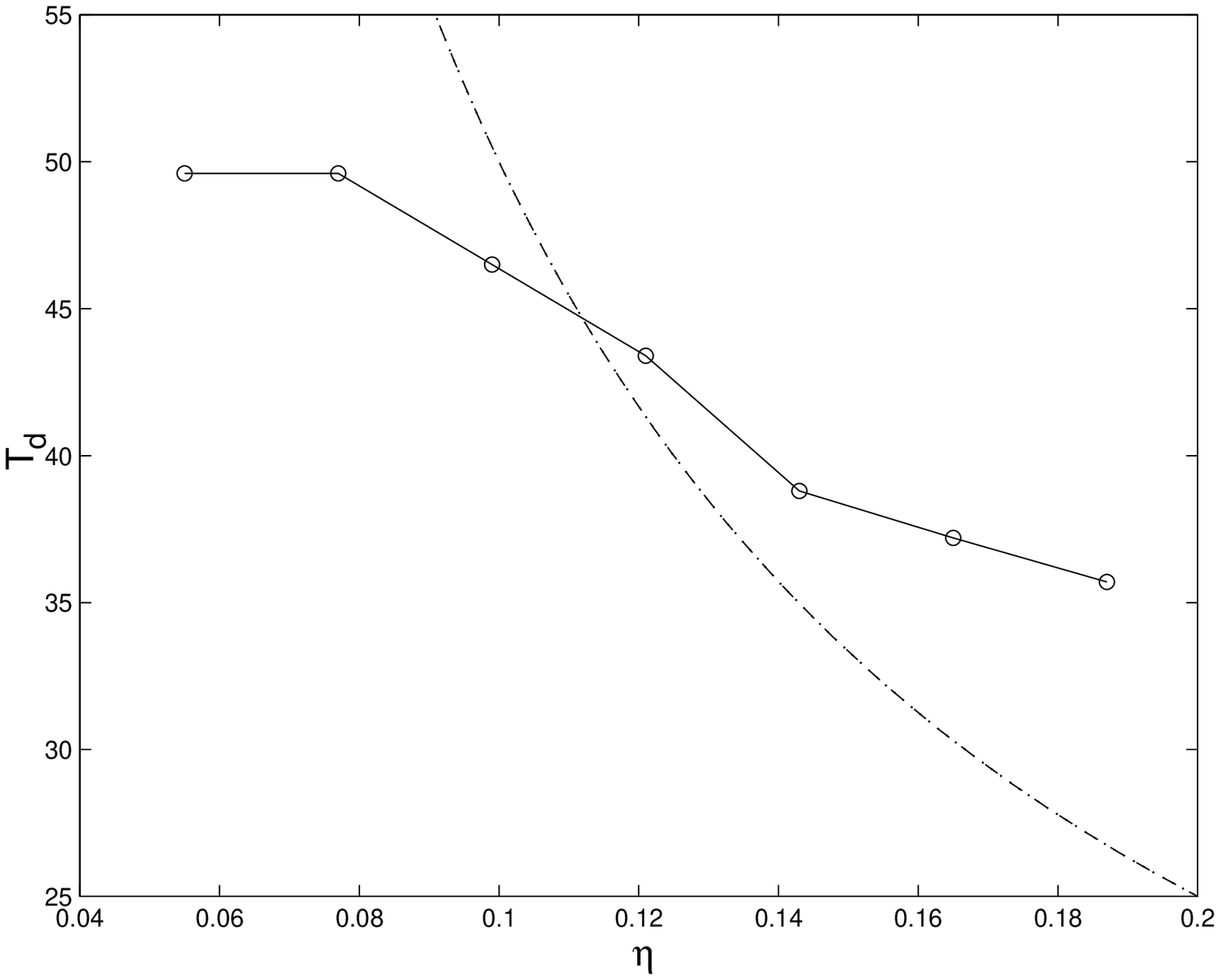}}
\caption{$T_d$ (in years) versus $\eta$ (in $10^{8}$ m$^2$ s$^{-1}$) for Model~1 ({\bf{Top Panel}}) and Model~2 ({\bf{Bottom Panel}}). The solid line connects the data points while the dashed line shows a C/x behavior. Other parameters are; $\alpha_0$ = 10 m s$^{-1}$, $v_0$ = 10 m s$^{-1}$, $B_0 = 10^5$ G, $B_c = 10^5$ G, $f_b = 0.05$ for Model~1 and $f_r = 0.06$ for Model~2.}
\end{figure} 
  
We show the variation of $T_d$ with the diffusivity $\eta$ for Model~1 (Top Panel) and Model~2 (Bottom Panel) in Figure~3. Again on comparison of the solid line connecting the data points with the dashed $\eta^{-1}$ line we find that the dynamo period is almost inversely proportional to the diffusivity within the SCZ for Model~1. The Bottom Panel presents the $T_d$ versus $\eta$ plot for Model~2, for $f_r = 0.06$ (corresponding to the {\it{buoyancy saturated regime}}). We find that in this case the dependence of $T_d$ on $\eta$ (the solid line) is far from a $\eta^{-1}$ dependence (the dashed line). In fact on re-doing the calculations for a higher value of $f_r$ for Model~2, we see that the dynamo period dependence on the diffusivity becomes less and less pronounced.
   
In PSKR models of the past with no meridional circulation and in interface dynamo models as well (Parker 1993; Markiel and Thomas 1999), an inverse dependence of $T_d$ on $\eta$ is expected and also seen. In simple linear models too, the period is expected to vary as $\eta^{-1}$. It is not a priori obvious that a non-linear model with $\alpha$ quenching and magnetic buoyancy - like Model~1, will have a $\eta^{-1}$ dependence of the dynamo period. Moreover, Dikpati and Charbonneau (1999) working with a BL type flux transport model with a similar {\it{recipe}} for buoyancy as our Model~2, reported a $T_d \propto \eta^{0.22}$ dependence. So the question naturally arises why does our Model~1 give such a strong $\eta^{-1}$ dependence at variance with other BL models.

In earlier studies we have presented some results of the variation in the latitude of eruptions with time (see for example Nandy and Choudhuri 2000 and Paper~I) for Model~1. We find that the strongest toroidal fields are usually found at high latitudes and their strength decreases progressively (due to eruptions) as they propagate towards the equator. Here, we find that with decreasing $\eta$ the region of eruptions start extending towards lower and lower latitudes. Presumably because with a low diffusivity, the strong toroidal fields can be stored for a longer time in the overshoot layer and get amplified by the strong radial shear (thus maintaining a value $>$ $B_c$) while it is being carried equatorward by the meridional flow. This increase in the region of magnetic activity increases the dynamo period with decreasing $\eta$ - simply because now the cycle has to extend and thus transport flux over a wider range in latitude. We shall see in Section~3.2 that indeed with decreasing $\eta$, stronger fields are found at lower latitudes within the overshoot layer. 
 
Another appealing reason may lie in the way the source term for the generation of the poloidal field is formulated. Note that due to the presence of the quenching term the poloidal field generation gets quenched when the erupted toroidal field approaches values close to or greater than $B_0$ ($10^5$ G). For a low value of $\eta$ and over a period of many successive eruptions, we find that the toroidal field near the top (where the $\alpha$-effect is concentrated) approaches very high values close to $B_0$. This is unacceptably large and quenches the poloidal field generation completely thus making the dynamo process inefficient. However when we increase $\eta$, the erupted field near the top diffuses and spreads out faster, thus not reaching very high values. This lets the poloidal field production go on uninterrupted making the dynamo process more effective. We believe that this scenario may also play a role in the reduction of the dynamo period with increasing $\eta$. 

As is clear from the above discussion, in Model~1, the poloidal field production is a two step process. With the toroidal flux first being transported to the top and then the $\alpha$-effect acting on it - with diffusion having the intermediate role of spreading out the erupted field. Contrary to that, in Model~2, the poloidal field production is a direct one step process, the efficiency of which is determined by $f_r$ and where the role of diffusivity is somewhat subdued. Therefore it is not surprising that with increasing $f_r$, the role of diffusivity in spreading out flux in Model~2 becomes more and more redundant and $T_d$ becomes less dependent on $\eta$. 
 
However, we never get a direct dependence of the period on diffusivity as reported by Dikpati and Charbonneau (1999) and this may be due to the differences that exists between our general model and theirs (for example they worked with a variable diffusivity profile which is such that variation of $\eta$ in the bulk of the SCZ does not affect the diffusivity within the overshoot layer much).   
  
\begin{figure}[t]
\centerline{\includegraphics[height=6.5cm, width=10cm]{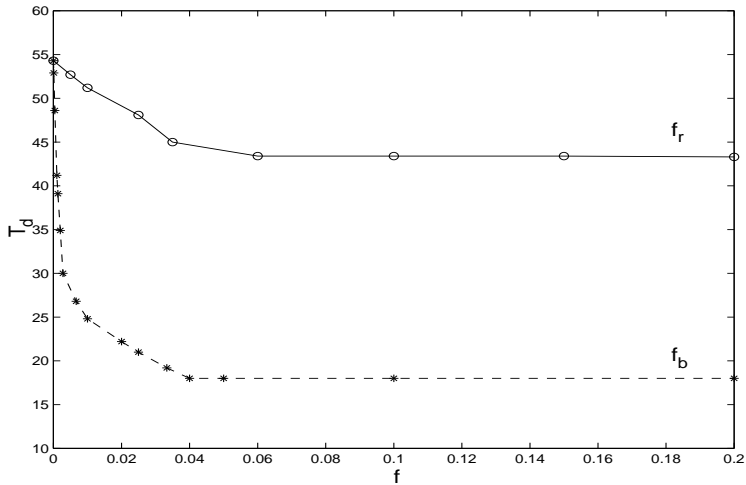}}
\caption{$T_d$ (in years) versus the fraction of the erupted field - $f_b$ for Model~1 (dashed line) and $f_r$ for Model~2 (solid line). Other parameters are; $\alpha_0$ = 10 m s$^{-1}$, $\eta = 0.12 \times 10^8$ m$^2$ s$^{-1}$, $v_0$ = 10.0 m s$^{-1}$, $B_0 = 10^5$ G, $B_c = 10^5$ G.}
\end{figure} 

Figure~4 shows the variation of the dynamo period with the control parameters $f_b$ (Model~1) and $f_r$ (Model~2), respectively. We had presented a similar plot for Model~1 in Paper~I (albeit with a lower value of $v_0$), we redo the calculation here for the sake of completeness of this paper and so as to easily compare it with Model~2. We see that $T_d$ for both the models decrease with increasing control parameter and saturates to a certain value - which is different for the two Models. As discussed in Paper~I, such a $T_d$ versus control parameter dependence characterises buoyancy driven flux transport models and portrays the fact that a higher control parameter means more efficient flux transport due to buoyancy and hence, a lower dynamo period. In that sense Model~2 (and the poloidal source formulation of Dikpati and Charbonneau 1999) does seem to capture the nature of buoyancy within a dynamo framework. We refer to the regime where the dynamo period has reached the saturation value, as the {\it{buoyancy saturated regime}}. For Model~1 this is found to occur at around $f_b = 0.04$ and for Model~2 this occurs at around $f_r =0.06$. 

Notice though that the period for Model~2 saturates at a much higher value. This essentially means that Model~2 is a less efficient manifestation of the buoyancy process. The depletion of the toroidal flux due to buoyancy in Model~1 plays a crucial role in limiting the latitudinal extent of the dynamo action (for a more detailed discussion on why this is so please refer to Paper~I). This in tandem with the efficient recycling of flux for a high $f_b$ decreases the dynamo period drastically.  However, Model~2 is formulated in such a way that it is not possible to deplete toroidal flux self-consistently. Thus the dynamo cycle takes place over the whole of the convection zone and there is hardly any effect on the toroidal field inside the overshoot layer by increasing $f_r$. Therefore it is not surprising that $T_d$ for Model~2 does not decrease as significantly as that of Model~1, with increasing control parameter. 

$B_c$ is constrained by results from simulations of flux tube rise and flux storage within the overshoot layer and is expected to be around $10^5$ G. Around a thousand eruptions (in the form of sunspots) is seen on the solar surface in a complete solar cycle and we have fixed the value of $\tau = 8.8 \times 10^5$ s to reflect that. However we did some runs with half and double the values of $B_c$ and $\tau$ and the dynamo period $T_d$ remained close to the original values. In any case in the {\it{buoyancy saturated regime}} for $f_b = 0.05$, $T_d$ is not expected to vary much with the parameters for buoyancy.

\subsection{The effect of buoyancy on the dynamo generated magnetic fields within the overshoot layer}

We have already seen from the results presented in the previous section that the strength of magnetic buoyancy ($f_b$) has a strong influence on the dynamo period (and thus on the efficiency) of the solar cycle. We carry this study of Model~1 further to see whether the mechanism of buoyant eruption has any effect on the magnitude of the magnetic fields inside the overshoot layer. 

The $\alpha$-quenching term can also constrain the magnitude of the dynamo generated magnetic fields. So it is necessary first to understand how this mechanism limits the magnetic field before we go on to the role played by buoyancy. As discussed in Section~2.1, the quenching term works in such a manner that poloidal field production stops rapidly once the erupted toroidal field approaches values close to $B_0 = 10^{5}$ G. This in turn has an effect on the magnitude of the toroidal field produced in the next cycle and over many cycles this mechanism ensures that the solutions converge to a stable oscillation with a non-growing amplitude of the magnetic fields. The results that we have presented are for such stable oscillations with a non-growing amplitude. Therefore the magnetic fields everywhere within the SCZ are expected to scale linearly with the value of $B_0$, if $\alpha$-quenching is the only magnitude limiting mechanism.
 
\begin{figure}[t]
{
\hspace{0.25cm}
\centerline{\includegraphics[height=6.5cm, width=10.8cm]{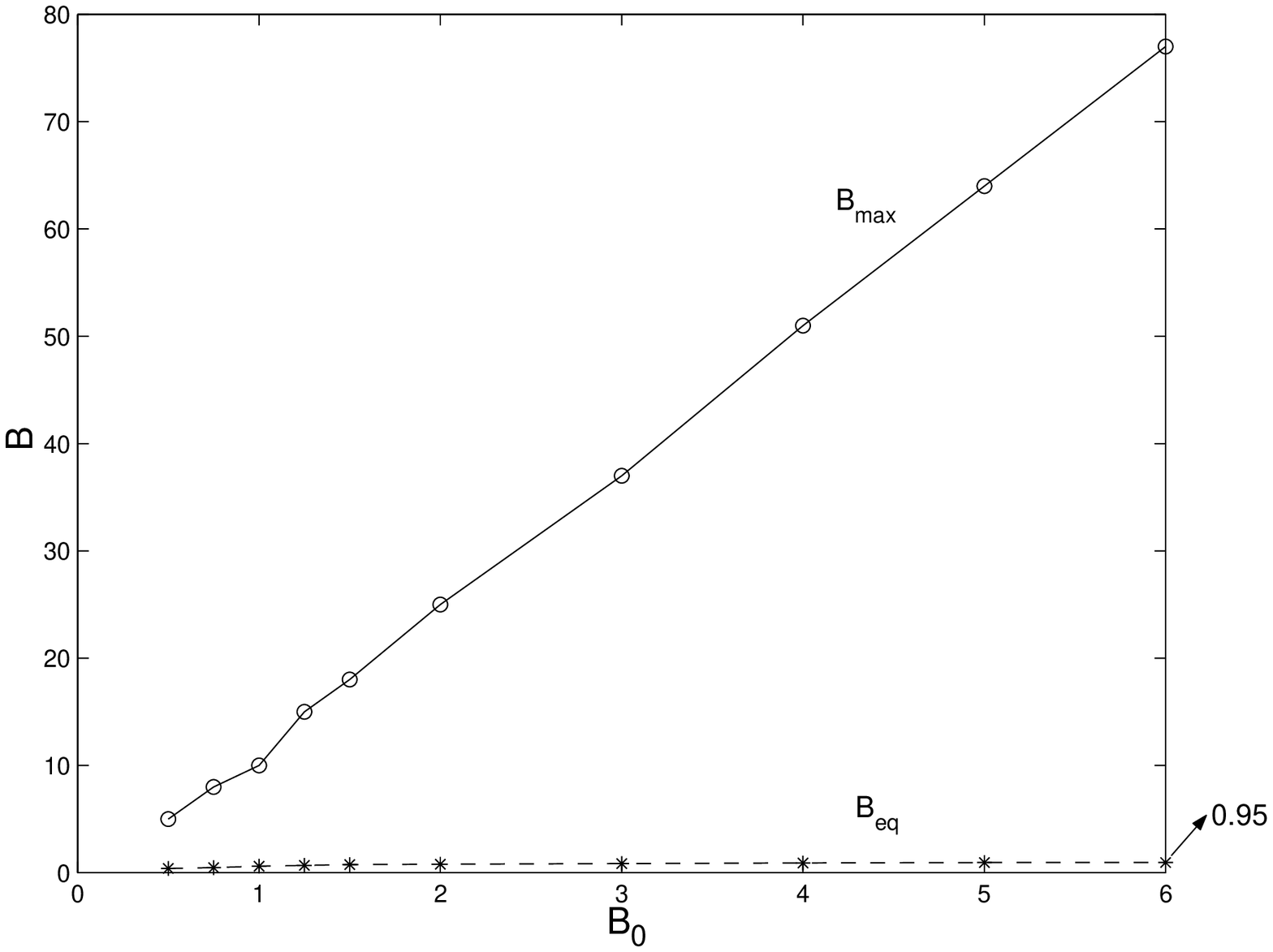}}
}
\centerline{\includegraphics[height=6.5cm, width=10cm]{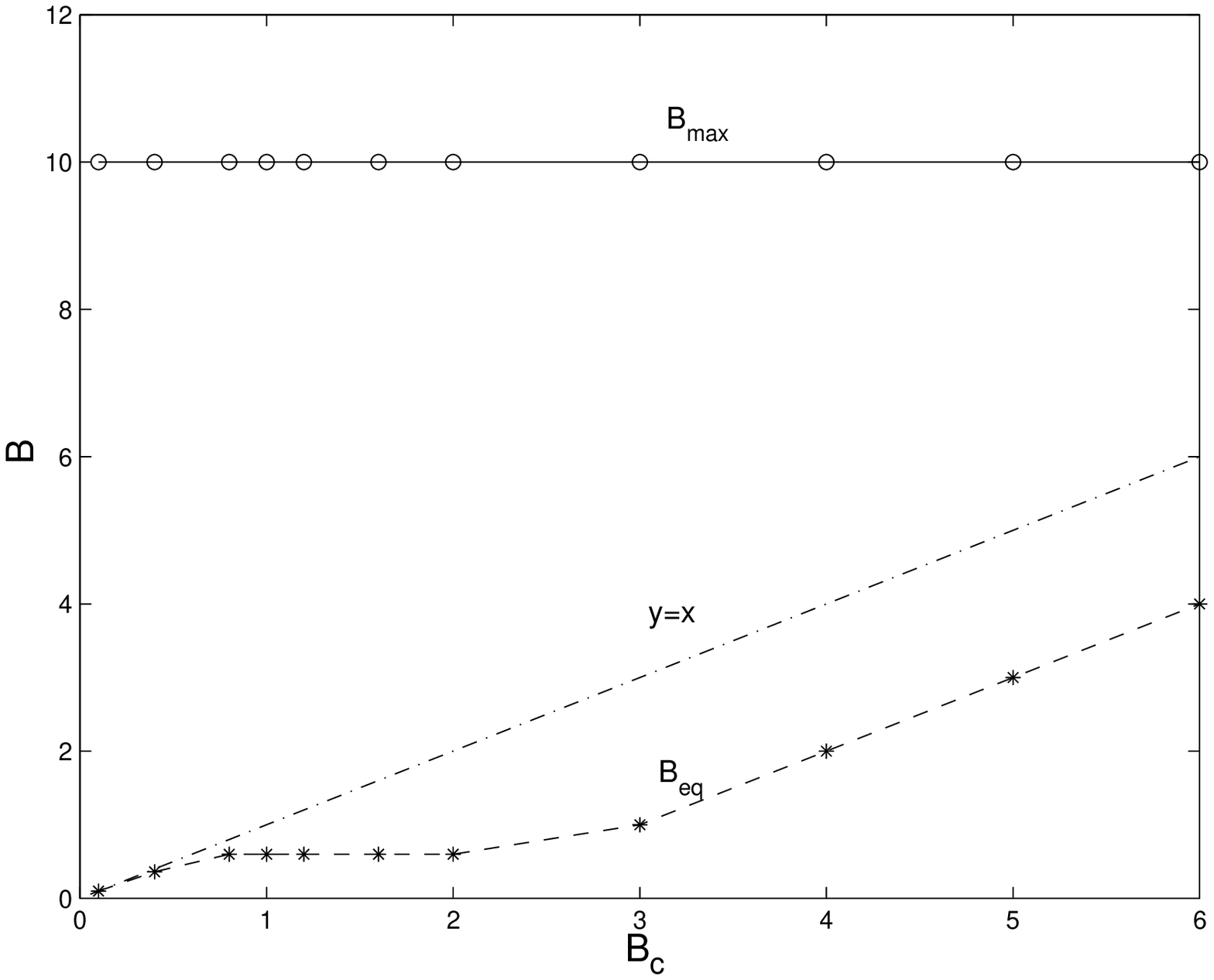}}
\caption{{\bf{Top Panel}}: Variation of the maximum toroidal field $B_{max}$ (solid line) and the low-latitude toroidal field $B_{eq}$ (dashed line) within the overshoot layer with the quenching field $B_0$, $B_c$ is fixed at $10^5$ G. {\bf{Bottom Panel}}: $B_{max}$ and $B_{eq}$ versus the critical field for eruption $B_c$, with $B_0 =10^5$ G. All fields are in units of $10^5$ G. The dash-dotted line in the Bottom Panel shows a y = x behavior for comparison. Other parameters are; $\alpha_0$ = 10 m s$^{-1}$, $\eta = 0.12 \times 10^8$ m$^2$ s$^{-1}$, $v_0$ = 10.0 m s$^{-1}$ and $f_b =0.05$.}
\end{figure} 

In Figure~5 (Top Panel) we present a plot of the variation of the toroidal fields within the overshoot layer with varying $B_0$, for $f_b = 0.05$. The solid line corresponds to the maximum toroidal field within the overshoot layer $B_{max}$ (which is found to be at high latitudes) and the dashed line corresponds to the toroidal field near the equator $B_{eq}$ at a latitude of around $10^0$. $B_{max}$ seems to be relatively unaffected by the adopted value for $B_c$ and scales with $B_0$, as expected for a model without buoyancy. Whereas $B_{eq}$ does not change much with $B_{0}$ and the maximum value it attains within the range that we have studied is $0.95 \times 10^{5}$ G. Now that is very significant, specially since the maximum value attained by $B_{eq}$ is {\it{slightly less than}} $B_c$. 

After the meridional down-flow drags the poloidal field down to the overshoot layer, the strong radial shear in the differential rotation starts working on it to create the toroidal field. By the time the toroidal field belt is advected down a little by the meridional circulation it reaches a high value and exceeds $B_c$ by about an order of magnitude. Eruptions start occurring immediately and as this toroidal field belt moves equatorward eruptions continue. Due to the accompanying depletion in field strength after eruptions, the toroidal field keeps on decreasing in strength until it falls below $B_c$ (obviously here the rate of flux production is less than the rate of flux depletion due to buoyancy). This therefore explains why the value of the toroidal field is constrained to $\leq$ $B_c$ at low latitudes. Just for a feeling for what would happen {\it{in the absence of buoyancy}}, consider the following; if for $B_0 = 10^{5}$ G $B_{eq}$ is found to be $10 \times 10^{5}$ G, on making $B_0 = 5 \times 10^{5}$ G, $B_{eq}$ attains a value of $ 50 \times 10^{5}$ G.
     
Figure~5 (Bottom Panel), where we present the variation of the toroidal field within the overshoot layer with the critical field $B_c$, lends further credence to the above inferences. We see that indeed $B_{max}$ remains unaffected by $B_c$ within the studied range, whereas $B_{eq}$ always stays below $B_c$, as is apparent on comparison with the dash-dotted $B = B_c$ line.

\begin{figure}[t]
\centerline{\includegraphics[height=6.5cm, width=10cm]{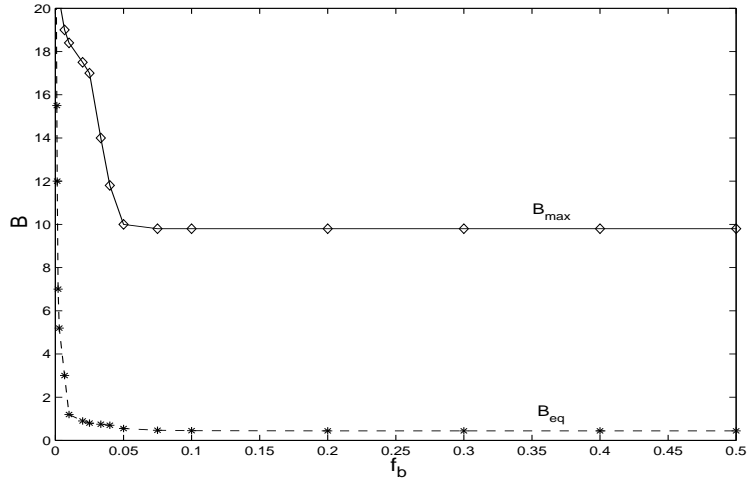}}
\caption{Variations in $B_{max}$ (solid line) and $B_{eq}$ (dashed line) with the fraction of the erupted field $f_b$. All fields are in units of $10^5$ G. Other parameters are; $\alpha_0$ = 10 m s$^{-1}$, $\eta = 0.12 \times 10^8$ m$^2$ s$^{-1}$, $v_0$ = 10.0 m s$^{-1}$, $B_0 = 10^5$ G, $B_c = 10^5$ G.}
\end{figure} 

We had done all the above calculations with a low value of $f_b = 0.05$ (though this value is in the {\it{buoyancy saturated regime}} of the dynamo). Even when such a low fraction of the toroidal flux is made to erupt we see that buoyancy manages to constrain the magnitude of the toroidal field. Naturally one wonders what would happen if we make the fraction of the erupted field (the control parameter) much larger. 

So we fix $B_c$ at $10^5$ G and study the variation of the toroidal field within the overshoot layer with $f_b$ in Figure~6. We see that with increasing $f_b$, both $B_{max}$ and $B_{eq}$ decreases and ultimately reaches an asymptotic limit. Interestingly, $B_{max}$ and $B_{eq}$ reaches their asymptotic limit at about the same value of $f_b$ for which the dynamo period reaches its asymptotic limit (see Figure~4). While $B_{max}$ drops to within a order of magnitude of $B_c$ (at $9.8 \times 10^5$ G), $B_{eq}$ drops down well below $B_c$ (at $0.44 \times 10^5$ G), thereby strengthening our conjecture that $B_{eq}$ is more affected by buoyant eruptions than $B_{max}$.  We may point out here that even with a higher $f_b$ (held constant), the previous result, that $B_{max}$ is not affected much by variation in $B_c$, holds true. 

We end this section by discussing a procedure, which may help us to fix an upper limit on the diffusivity within the solar overshoot layer. This technique is motivated from an understanding of the results presented above. Briefly summarising the results relevant to this analysis, we have learned that when a strong toroidal field belt inside the overshoot layer is advected equatorward by the meridional circulation, it decreases in strength due to buoyant eruptions. Careful study of the variation in the eruption latitude shows us that with decreasing diffusivity the region of eruption extends to lower and lower latitude. This led us to hypothesise (in Section~3.1) that with a lower value of diffusivity within the overshoot layer, it may be possible to store and amplify the toroidal field belt so that it is maintained above the critical field $B_c$ even at low latitudes, hence allowing for eruptions there, as seen in reality. 

\begin{figure}[t]
\centerline{\includegraphics[height=6.5cm, width=10cm]{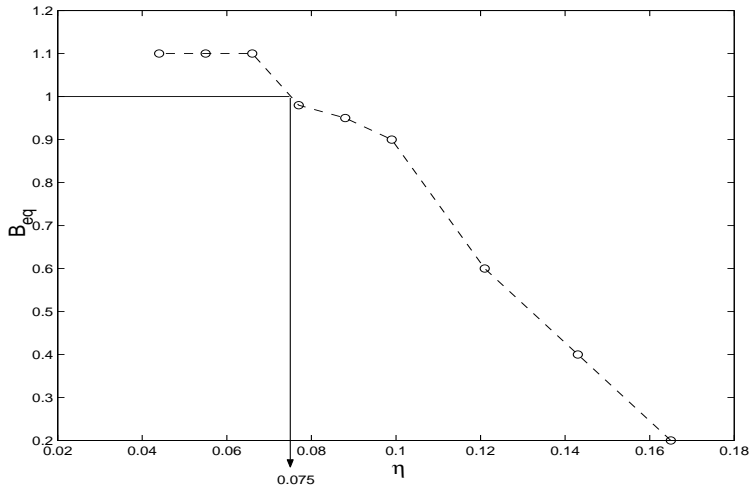}}
\caption{Variation of the low-latitude toroidal field $B_{eq}$ within the overshoot layer, with the diffusivity $\eta$. $B_{eq}$ is in $10^5$ G and $\eta$ is in $10^8$ m$^2$ s$^{-1}$. The dashed line shows this variation while the solid line is the intercept corresponding to $B_{eq} = B_c$. Other parameters are; $\alpha_0$ = 10 m s$^{-1}$, $v_0$ = 10 m s$^{-1}$, $B_0 = 10^5$ G, $B_c = 10^5$ G, $f_b = 0.05$.}
\end{figure} 

In Figure~7 we plot the variation of the toroidal field within the overshoot layer near the equator (at $10^0$ latitude) with respect to the diffusivity $\eta$. The dashed line shows this variation and the result that the strength of the low-latitude toroidal field inside the overshoot layer falls with increasing $\eta$, lays a more solid foundation to our starting hypothesis. We find that on making $\eta > 0.075 \times 10^{8}$ m$^2$ s$^{-1}$, the strength of the toroidal field within the overshoot layer falls below $B_c = 10^5$ G. This suggests that the upper limit of the diffusivity within the overshoot layer (which we may call $\eta^{max}_{overshoot}$) should be $0.075 \times 10^{8}$ m$^2$ s$^{-1}$, to ensure eruptions at low latitudes as seen in reality. Note that the value of the diffusivity also depends on the adopted thickness of the overshoot layer.

However, on the basis of this model alone, we cannot make a claim to the authenticity of the value of $\eta^{max}_{overshoot}$ as found above. Rather we have shown that based on some physical arguments, {\it{it is possible}} to make such an indirect estimate (without taking into account the effects that a strong magnetic field may have on $\eta$ inside the overshoot layer). The exact value of $\eta$ within the overshoot layer remains to be verified by other independent analysis, preferably from a more fine tuned solar dynamo model.

\section{Concluding remarks}

The basic foundation of this model was laid in Paper~I, where we showed that a model with such a {\it{recipe}} for buoyancy and with a concentrated $\alpha$-effect near the surface, is a valid representation of the BL approach. In this paper, we have carried this study further, to show how buoyancy affects the dynamo generated magnetic fields and the working of the solar dynamo in general.     
 
The strength of magnetic buoyancy is seen to affect the dynamo period drastically. With increasing fraction of the field which is taken up $(f_b)$, the dynamo period decreases and reaches an asymptotic limit - the {\it{buoyancy saturated regime}} of the dynamo. One important question is ofcourse whether the solar dynamo is actually working in this {\it{regime}} or whether it is working at a {\it{non - buoyancy saturated regime}}. In the former case, variation in the parameters controlling buoyancy will not have much effect on the dynamo period (and the dynamo generated magnetic fields) and it will be the meridional flow speed which will primarily control the period, whereas in the latter case, variations in the control parameters for buoyancy will strongly influence the dynamo period (and also the amplitude of the cycle). 

Since we find that for a very low value of $f_b = 0.05$ the dynamo period reaches its asymptotic limit, chances are fairly high that the solar dynamo is indeed working at the {\it{buoyancy saturated regime}} and meridional flow speed and its fluctuations has the final say in determining the cycle period, thus acting as a {\it{solar clock}}. Some authors have studied the effects of stochastic fluctuations in BL models of the solar dynamo and their possible effects on the cycle period and amplitude, see for example Charbonneau and Dikpati (2000) and Charbonneau (2001, in press). A strong influence of the meridional circulation on the dynamo cycle period and amplitude is portrayed in their results. In Section~3.1, following Figure~1 we surmised about the crucial role played by the meridional down-flow near the poles in completing the dynamo chain and meridional flow being the slowest process in this chain, in all likelihood {\it{is}} the main determinant of the solar cycle period.

In retrospect, it is surprising that the dynamo saturates and reaches the {\it{buoyancy saturated regime}} at such a low fraction of the erupted field. Note however that when flux tubes become buoyant and start to rise, gravity would stretch the field lines (due to the rapid rise of the upper lighter part of the tube) and the field reconnects. It is not inconceivable than that the lower part of the reconnected tube (with the larger fraction of the flux) which is rooted to the overshoot layer, sinks back into the overshoot layer. Some studies also suggest that a large fraction of the erupted flux may actually be retracted back into the deeper layers of the SCZ (Rabin, Moore, and Hagyard 1984; Parker 1984, 1987, Howard 1992; D'Silva 1995) thus not contributing to the poloidal field regeneration. These considerations lead us to conclude then that maybe in reality, it is indeed a small fraction of the deep toroidal field, which contributes to the flux recycling process.    
   
Within the framework of such a buoyancy driven flux transport model, we find that the diffusivity $\eta$ and its relative magnitude in the lower and upper parts of the SCZ is of vital importance, even though its role as a flux-transporter between the two source regions is greatly undermined by that of the meridional flow and magnetic buoyancy. While a low value of the diffusivity is required within the overshoot layer, to enable toroidal fields exceeding the critical field limit $B_c$ to be present at low latitudes (thus resulting in eruptions there), a higher value of diffusivity may make the poloidal flux generation near the surface more efficient by spreading out the erupted toroidal field (so that the $\alpha$-effect is not quenched). This latter scenario remains to be explored more quantitatively (with a depth-dependent diffusivity) and a study of the same will be undertaken in the near future.    
  
Most solar theorists seem to agree on the value of $10^8$ m$^2$ s$^{-1}$ as an upper limit for $\eta$ for the convection zone proper and surface observational estimates also point to a similar figure (Wang, Nash, and Sheeley 1989a,b; Dikpati and Choudhuri 1995; Schrijver and Martin 1990). With respect to the above figure, some theoretical arguments can be made to make an order of magnitude estimate of the value of $\eta$ within the overshoot layer (Parker 1993), where the magnetic field is an order of magnitude greater than the magnetic field within the SCZ. Following these arguments it turns out that $\eta$ inside the overshoot layer should be two orders of magnitude less than $\eta$ in the main body of the SCZ. That is, $\eta$ should be around $0.01 \times 10^8$ m$^2$ s$^{-1}$ inside the overshoot layer. We have proposed a mechanism for estimating an upper limit of the diffusivity within the overshoot layer and have come up with a value, $\eta^{max}_{overshoot} = 0.075 \times 10^{8}$ m$^2$ s$^{-1}$. Though this result has been arrived at with a rather simple dynamo model (with only a radial shear in the rotation), it is nice to see that it does not contradict the earlier speculative value.
 
We have shown that magnetic buoyancy can limit the magnetic field within the overshoot layer and the adopted value for the critical field $B_c$ strongly constrains the toroidal field at low latitudes. One question naturally arises here - is magnetic buoyancy capable of quenching the growth of the dynamo within the framework of such models? We did some runs with infinite $B_0$ (that is no $\alpha$ quenching) and with $B_c = 10^5$ G to test this. In this case, we found that the amplitude of the generated fields kept on blowing up without saturating to a finite-amplitude oscillation (note that with an infinite $B_0$, the equations become linear once the toroidal field exceeds $B_c$ and hence the result that the generated fields blow up, is a necessary outcome). Thus the answer to the above question is - no. At least, within the framework of such kinematic dynamos, where there is an {\it{infinite}} energy source to tap from (the prescribed meridional motions and the differential rotation), magnetic buoyancy alone, is not capable of quenching the growth of the dynamo generated fields. Thus magnetic buoyancy seems to limit the dynamo generated fields {\it{within}} a larger quenching framework.     

Our results also show that the peak deep toroidal field attains a very high value within a short span of the onset of a new half-cycle and after that the toroidal field strength continously decreases till the end of that half-cycle, the pattern repeating itself. Assuming that the sunspot activity is directly related to the toroidal field at the bottom of the SCZ (we may point out here that it is still not clear how strong flux tubes form out of a diffuse field and whether the strength of sunspots reflect the strength of the toroidal flux tubes in the overshoot layer), this would translate to seeing the strongest active regions within a couple of years of the beginning of a new cycle and relatively weaker and weaker active regions as the sunspot cycle progresses (at lower latitudes). At the end of 11 years this cycle would repeat itself. There exists observational evidence which shows that sunspot strength and size is maximum during the solar maximum (around 5.5 years after the start of a new cycle) and decreases progressively till the minimum (Tang, Howard, and Adkins 1984). This is reflected to an extent in the presented results albeit with a higher latitude belt of activity and with an offset of a couple of years. It remains to be seen whether a more sophisticated model incorporating the latitudinal dependence of the differential rotation can reproduce the observations exactly.

We end with a few critical comments on the rather simple model that we have used. Our model incorporates a concentrated radial shear at the base of the SCZ matching the helioseismologically determined profile from mid-latitudes to low latitude near the equator and is deficient in the sense that we have not considered the latitudinal variation in the rotation. We believe that the inclusion of a latitudinal shear will not change the results of the parameter space study qualitatively. Nor is it going to change the result that buoyant eruptions, limits the toroidal field inside the overshoot layer, which is a fundamental outcome of the eruption and subsequent flux depletion procedure. What it may change though is the nature and appearance of the magnetic butterfly diagrams. Due to this we have refrained from going for any detailed comparison with observations in this paper.

A latitude-dependent solar like rotation profile might also influence at which latitude we find the maximum toroidal fields. However, Durney (1997) and K\"uker, R\"udiger, and Schultz (2001) working with a solar like rotation profile, also finds the maximum toroidal field at high latitudes near the pole. While these results support our case, we cannot help but acknowledge that the finding of the maximum toroidal fields at high latitudes is a definite problem which needs to be addressed at some stage. That however is beyond the scope of this model and we leave it to future studies to address this issue. 

\acknowledgements
   
I would like to thank Paul Charbonneau and Bernard Durney, lively exchanges with whom inspired some of the studies and discussion presented in this paper. I am also grateful to Arnab Rai Choudhuri, initial work with whom laid the foundations of the model explored in this study.

\pagebreak

\end{article}
\end{document}